
\voffset=3truecm
\hoffset=2.25truecm
\baselineskip=16pt
\rightline{ILL-(TH)-93-21}
\vskip1truecm

\centerline {\bf  COMPOSITENESS, TRIVIALITY AND BOUNDS ON CRITICAL}
\centerline  {\bf EXPONENTS FOR FERMIONS AND MAGNETS }
\vskip1truecm

\centerline {Aleksandar KOCI\' C and John KOGUT}
\centerline {\it Loomis Laboratory of Physics, University of Illinois,
                 Urbana, Il 61801}
\vskip3truecm

\centerline {\bf Abstract}
\vskip 0.3truecm

We argue that theories with fundamental fermions which undergo chiral
symmetry breaking have several universal features which are qualitatively
different than those of theories with fundamental scalars. Several bounds
on the critical indices $\delta$ and $\eta$ follow. We observe that
in four dimensions the
logarithmic scaling violations enter into the Equation of State of
scalar theories, such as
$\lambda\phi^4$, and fermionic models, such as
Nambu-Jona-Lasinio, in qualitatively different ways.
These observations lead to useful approaches for analyzing lattice
simulations of a wide class of model field theories. Our results imply that
$\lambda\phi^4$ {\it cannot} be a good guide to understanding the
possible triviality of spinor $QED$.

\vfill\eject

\noindent {\bf 1. Introduction}
\vskip 0.3truecm

There are two classes of theories in the literature
that are used to model the Higgs sector of the Standard Model [1,2].
One is based on the self-interacting $\phi^4$ scalar theory in which
the Higgs is elementary.  The other is based on strongly interacting
constituent fermi fields in which the Higg's particle is a
fermion-antifermion bound state.
A recent proposal for the second realization
uses Nambu-Jona-Lasinio(NJL) models in which composite scalars emerge
as a consequence of spontaneous chiral symmetry breaking [2].
In four dimensions both types of models have a trivial
continuum limit and are meaningful only as effective theories with a
finite cutoff. This restriction places constraints on the low-energy
parameters e.g. bounds on the masses. We wish to point out in this letter
that triviality in the two models is realized in {\it different}
ways. The differences between theories of composite and elementary mesons
can be expressed in terms of the critical indices $\delta$ and $\eta$, and
several inequalities and bounds on these indices will follow. These results
should prove useful in theoretical, phenomenological and
lattice simulation studies of a wide class of model field theories.

We begin with a few comments about the physics in each model.
In a NJL model [3], as a result of spontaneous
chiral symmetry breaking, the pion-fermion coupling is
given by the Goldberger-Treiman relation $g_{\pi N}=M_N/f_\pi$, where $M_N$
is the fermion mass and $f_\pi$ is the pion decay constant. Being the
wavefunction of the pion, $f_\pi$ determines its  radius as well:
$r_\pi \sim 1/f_\pi$. The Goldberger-Treiman can then be written in the
suggestive form $g_{\pi N}\sim M_N r_\pi$. Thus, the coupling
between pions and fermions  vanishes as the size of the pion shrinks to zero.
The origin of triviality of the
Nambu-Jona-Lasinio model is precisely the {\it loss} of compositeness of the
mesons [3]. The force between the fermions is so strong that the constituents
collapse onto one another producing pointlike mesons
and a noninteracting continuum theory. In a self-interacting scalar theory,
like $\phi^4$, the mesons are elementary and the reason for triviality
is different [4,5]. At short distances the interaction is repulsive, so there
is no collapse. The structure of the scalars, needed for the interaction to
survive
the continuum limit, should be built by weakly interacting bosons.
In four dimensions, the short ranged $\lambda\phi^4$
interaction fails  to provide a
physical size for the mesons. It cannot be felt
by the particles because of the short-distance repulsion -- they
cannot meet where collective behavior can set in
and produce macroscopic fluctuations. In this way
the cutoff remains the only scale and the continuum limit is trivial.

Consider two simple, soluble examples:
the large-$N$ limits of
the $O(N)$ $\sigma$- and the four-fermi model [6,7].
They exhibit a phase transition at finite coupling for $2<d<4$.
Their critical exponents are given in Table 1.
As is apparent from the Table, the two sets of critical
indices evolve differently when $d$ is reduced below 4.
At first glance this might be surprising since both models break the
same symmetry spontaneously and one expects that
they describe the same low energy physics.
The purpose of this paper is to show that this difference between
the critical exponents is generally
valid, irrespective of the approximations employed.
As a consequence of this it will be possible to establish
a bound on the exponent $\delta$ which for scalar theories
is $\delta\geq 3$, and for fermionic theories is $\delta\leq 3$.
Although in four dimensions the two sets of exponents coincide,
they are accompanied by logarithmic corrections due to
scaling violations.
It will be shown that the bounds announced above are respected by the
scaling violations as well, and that consequently
the scaling violations have {\it opposite} signs in the two classes of
theories.
These bounds are a consequence of different realizations
of symmetry breaking, the essential difference being
the fact that for scalar theories mesons are elementary, while in the
case of the chiral transition in fermionic theories, they are composite.
The bounds on $\delta$ are just another way of expressing this difference
in terms of universal quantities. Finally, we
will discuss the implications of these results on
triviality in both models in four dimensions.
\baselineskip=20pt
\centerline{\bf Table 1}
\vskip 0.5 truecm
\centerline
{Leading order critical exponents for the spherical and four-fermi model}
\vskip 0.5 truecm
$$\vbox{\settabs\+\qquad18\qquad&\qquad190.4(8.8)\qquad&\qquad.1630(53)
\qquad&\cr
\+\hfill exponent\hfill&\hfill$\sigma$-model\hfill&\hfill
four-fermi\hfill&\cr\bigskip
\+\hfill$\beta$\hfill&\hfill${1\over
2}$\hfill&\hfill${1\over{d-2}}$\hfill&\cr\smallskip
\+\hfill$\nu$\hfill&\hfill${1\over{d-2}}$\hfill&\hfill${1\over{d-2}}$\hfill&\cr\smallskip
\+\hfill$\delta$\hfill&\hfill${{d+2}\over{d-2}}$\hfill&\hfill$d-1$\hfill&\cr\smallskip
\+\hfill$\gamma$\hfill&\hfill${2\over{d-2}}$\hfill&\hfill$1$\hfill&\cr\smallskip
\+\hfill$\eta$\hfill&\hfill$0$\hfill&\hfill$4-d$\hfill&\cr\smallskip
\+\qquad$\,$\qquad&\qquad$\,$\qquad&\qquad$\,$\hfill&\cr}$$
\vskip0.5truecm
\baselineskip=16pt

\noindent {\bf 2. Mass ratios and bounds on $\delta$}
\vskip0.3truecm
To approach the problem, it is convenient to adopt a particular view of
the phase transition [8].
Instead of the order parameter we will use
mass ratios to distinguish the two phases. While the order
parameter is a useful quantity to parametrize the phase diagram,
the spectrum carries direct information
about the response of the system in it's different phases and its form does not
change in the presence of an external symmetry-breaking field.
In what follows we will switch from magnetic to chiral notation
without notice. The correspondence is: magnetic field ($h$)
$\Leftrightarrow$ bare mass ($m$); magnetization ($M$) $\Leftrightarrow$
chiral condensate
($<\bar\psi\psi>$); longitudinal and transverse modes $\Leftrightarrow$
$(\sigma$, $\pi)$; $h\to 0$ $\Leftrightarrow$ chiral limit.
Theories that treat scalars
as elementary will be referred to as 'magnets' and those that give rise to
composite mesons as a consequence of spontaneous chiral symmetry breaking
will be refered to as 'fermions'.

Consider the effect of spontaneous symmetry breaking on the spectrum
from a physical point of view. In the symmetric phase,
there is no preferred direction and symmetry
requires the degeneracy between longitudinal and transverse modes
(chiral partners). Therefore, in the zero-field (chiral) limit
the ratio $R=M_T^2/M_L^2=M^2_\pi/M^2_\sigma \to 1$.
As the magnetic field (bare mass) increases, the ratio
decreases (because of level ordering, $\sigma$ is always heavier then $\pi$).
In the broken phase, however, the ratio vanishes in the chiral
limit because the pion is a Goldstone boson. This time, the ratio clearly
increases away from the chiral limit. The higher the magnetic field,
the less important is the manner in which
the symmetry is effected
by the dynamics. The value of the ratio at large $h$
is less sensitive to variations in the coupling.
The qualitative behavior of the mass ratio
is sketched in Fig.1.
The important property of the mass ratio, in this context, is that
its properties follow completely from the properties of the order
parameter [8]. This, after all, comes as no surprise since
both quantities, $M$ and $R$, contain the same physics
and merely reflect two aspects of one phenomenon.

The essential ingredients are the Equation of State (EOS)
and the Ward identity which
follows from it.

$$
h_a=M_a M^{\delta-1}f\biggl(t/M^{1/\beta}\biggr),\,\,\,\,\,\,\,\,
\chi_T^{-1}=h/M, \,\,\,\,\,\,\,\, \chi_L^{-1}={ {\partial h}\over
{\partial M} }
\eqno(1)
$$
where $t$ is the usual reduced coupling (temperature).
The second equation follows from the first upon differentiation. It
is the Ward identity that ensures the Goldstone theorem.
The ratio of susceptibilities is a function of the reduced variable,
$t/M^{1/\beta}$,
only and is determined completely by the critical indices and the
universal function $f(x)$. Elementary algebra shows that the ratio
$\chi_L/\chi_T$ is
a function of one variable only, namely $R(t,h)=R(h/t^\Delta)$ which explains
the behavior suggested by Fig.1.
At the critical point the order parameter scales as $M\sim h^{1/\delta}$
The susceptibility ratio is just a
logarithmic derivative, $R=\partial\ln M/\partial\ln h$.

$$
{1\over R}={ {\chi^{-1}_L}\over{\chi^{-1}_T} }=\delta, \,\,\,\,\,\,\,\, t=0
\eqno(2)
$$
At the critical point it is independent of the symmetry breaking field.
The exponent $\delta$ measures the relative strength
of the transverse and longitudinal responses [8].

The plot of $R$ versus $h$ (Fig.1) with the critical
isotherm is a "universal phase diagram" --
any curve $R>1/\delta$ belongs to the symmetric phase and
$R<1/\delta$ to the broken phase.
In this way,
the entire range of coupling constant maps
onto the interval $[0,1]$. The critical coupling
maps onto $1/\delta$. In these units
the size of the broken phase is $1/\delta$ and of the symmetric phase
$1-1/\delta$.

To explain the relevance of these observations to the problem in question,
we recall
some facts about magnets and compare them to the chiral transition.
\vskip3truemm

\noindent
{\it -- Magnets}:
At zero magnetic field, the critical coupling (temperature) separates two
regions: the low-temperature (weak coupling) from the
high-temperature (strong coupling) phase.
Spontaneous symmetry breaking occurs at low temperatures.
As the dimensionality is reduced towards the lower critical value
($d=2$), the disordering effect of infrared fluctuations becomes more important
and the broken (massless) phase shrinks. On the "universal phase
diagram", Fig.2a, this implies that $1/\delta$ decreases below its mean field
value. Thus, in this case $\delta>3$.
\vskip3truemm

\noindent
{\it -- Fermions}: For the chiral transition the same reasoning applies
except
that now the {\it broken and symmetric phases are interchanged}. Chiral
symmetry breaking is a strong coupling phenomenon. Thus, in this case
it is the symmetric phase that shrinks in lower dimensions. As is seen
in Fig.2b, this implies that $1/\delta$ increases, leading to
$\delta<3$.

These are the bounds announced in the introduction.
\vskip5truemm

\noindent{\bf 3. Scaling violations and triviality in four dimensions}
\vskip3truemm
One way to determine the sign of the scaling violations in four dimensions
is to proceed in the spirit of the $\epsilon$-expansion i.e.
to approach four dimensions from below [9]. The transcription to the language
of
scaling violations is established by the replacement
$\epsilon\to 1/\log\Lambda$ in the limit $\epsilon\to 0$. Thus, the extension
of the arguments made before for $d<4$ can be made by simply taking
the limit $d\to 4$. In this way we anticipate that
the two inequalities prevail and suggest
that the scaling violations have different signs in the two theories.

The difference in the sign of the scaling violations
in fermions and magnets has a simple explanation and
lies at the root of the difference between the patterns of symmetry
breaking in the
two systems. Imagine that we fix the temperature to its
critical value $T=T_c$ and approach the critical point $(T=T_c, h=0)$ in the
$(T,h)$ plane from the large-$h$ region.
The possible similarity of the two models is related to their
symmetry. This is apparent in the chiral (zero field) limit where this
symmetry is manifest. By going away from this limit chiral symmetry is
violated and the two models differ.

Consider the behavior of the mass ratio for magnets in a
strong magnetic field, away from the scaling
region. In this regime, the temperature
factor can be neglected and the hamiltonian describes free
spins in an external field ($H\to h\sum_i S_i$).
The energy of longitudinal excitations is
proportional to the field-squared, $\chi_L^{-1}\sim h^2$,
while the transverse mass  remains
fixed by the symmetry, namely $\chi_T^{-1}=h/M$ for any value of $h$.
The effect of the external field is to introduce a preferred direction and its
increase results in amplification of the difference between the
longitudinal and transverse dirrections. For large
$h$ the ratio scales as
$R\sim 1/h$. Therefore, an increase in magnetic field
reduces the ratio towards zero. The critical isotherm in this case
bends down (Fig.3).

For fermions, the mesons are
fermion-antifermion composites. Close to the chiral limit, they are collective.
However, as the constituent mass increases, they turn into atomic states
and the main contribution to the meson mass comes from the rest energy
of its constituents. In the limit of infinite bare mass,
interactions are negligible and $M\to 2m$ regardless of the channel. Thus,
outside of the scaling region, an increase in $m$ drives the ratio to 1
(Fig.3).

{\it Thus, the scaling violations for magnetis and fermionis have
opposite signs}. They contain knowledge of the physics away from
the chiral limit where the two models are quite different and these
differences remain as small corrections close to the chiral limit.

To establish the connection between scaling violations and triviality,
we introduce the renormalized coupling. It is a dimensionless low-energy
quantity
that contains information
about the non-gaussian character of the theory. It is conventionally
defined as [10]

$$
g_R=- { {\chi^{(nl)}}\over{\chi^2 \xi^d} }
\eqno(3)
$$
where the nonlinear susceptibility $\chi^{(nl)}$ is the zero-momentum
projection
of the connected four-point function

$$
\chi^{(nl)}={ {\partial^3 M}\over{\partial h^3} }=
\int_{123}<\phi(0)\phi(1)\phi(2)\phi(3)>_c
\eqno(4)
$$
The normalization fators, $\chi = \int_x<\phi(0)\phi(x)>_c$ and $\xi^d$,
in eq.(3) take
care of the four fields and the three integrations.
In a gaussian theory all higher-point functions
factorize, so $g_R$ vanishes.
Using the hyperscaling hypothesis, this can be converted into

$$
g_R \sim \xi^{(2\Delta-\gamma-d\nu)/\nu}
\eqno(5)
$$
where $\Delta=\beta+\gamma$ is the gap exponent.
Being dimensionless, $g_R$ should be independent of $\xi$ if $\xi$ is the only
scale. Thus, the validity of hyperscaling
requires that the exponent must vanish. It
implies the relation, $2\Delta-\gamma -d\nu=0$,
between the critical indices.
In general, it is known that the following inequality [11] holds

$$
2\Delta\leq \gamma+d\nu
\eqno(6)
$$
The exponent  in the expression for $g_R$ is always
non-positive, so that violations of hyperscaling imply that the resulting
theory  is
non-interacting.

Above four dimensions, the exponents are gaussian ($\gamma=1,\Delta=3/2,
\nu=1/2$). In this case, it is easy to verify the above inequality:
$3\leq 1+d/2$, which amounts
to $d\geq 4$. In four dimensions most field theoretical models
have mean-field critical exponents, but
with logarithmic corrections that drive $g_R$ to
zero.
Scaling violations in any thermodynamic quantity propagate into the
renormalized
coupling and, according to eq.(6), these violations lead
to triviality.

Instead of using the ideas of the $\epsilon$-expansion where
scaling is always respected and where equalities between
exponents hold, we will fix $d=4$ and compute the logarithmic corrections
to the critical exponents.
In order to focus on the problem in question, we
analyze two simple models:
$\phi^4$ and $(\bar\psi\psi)^2$ theories both in the
large-$N$ limit. The results that will be discussed are completely
general and the two models are chosen just to make the argument simple.
The effective actions for the two models are [12,13]

$$
V(M)=-{1\over 2}tM^2+{\lambda\over 4} {{M^4}\over{\log(1/M)}}
\eqno(7a)
$$
$$
V(<\bar\psi\psi>)=-{1\over 2}t<\bar\psi\psi>^2+
<\bar\psi\psi>^4\log(1/<\bar\psi\psi>)
\eqno(7b)
$$
This is the leading log contribution only.
In the first example, it is clear how log-corrections lead to
triviality. The logarithm can simply be thought of as coming from the
running coupling --
quantum corrections lead to the replacement $\lambda\to \lambda_R$.
The vanishing of the renormalized coupling is then manifest from eq.(7a).
In the case of fermions, eq.(7b),
the details are completely different -- the analogous
reasoning would lead to an erroneous
conclusion that the renormalized coupling increases in the infrared. In
eq.(7b) the explicit coupling is absent from
the fluctuating term -- it is already absorbed in the curvature. Once the
curvature is fixed, the effective coupling is independent of the bare one.
The vanishing of the renormalized coupling here follows
from the wave funciton renormalization constant
$Z\sim 1/\ln(1/<\bar\psi\psi>)$ [13].

In both cases the renormalized coupling is obtained through the
nonlinear susceptibility. For simplicity, we work in the symmetric phase
where the odd-point functions vanish. The correlation length is related to the
susceptibility by $\xi^2=\chi/Z$. For magnets the folowing relations hold

$$
\chi^{(nl)}\sim \chi^4 {\lambda\over{\ln(1/M)}},\,\,\,\,\,\,\,\,\,\,\,\,
Z=1
\eqno(8a)
$$
$$
g_R\sim Z^2 {\lambda\over{\ln(1/M)}}\sim {1\over{\ln(1/M)}}
\eqno(8b)
$$
so $g_R$ vanishes at a logarithmic rate.

For fermions, on the other hand, we have

$$
\chi^{(nl)}\sim \chi^4 \ln(1/<\bar\psi\psi>),\,\,\,\,\,\,\,\,\,\,\,\,
Z\sim{1\over{\ln(1/<\bar\psi\psi>)}}
\eqno(9a)
$$
$$
g_R\sim Z^2 \ln(1/<\bar\psi\psi>)\sim {1\over{\ln(1/<\bar\psi\psi>)}}
\eqno(9b)
$$
In this context the following point should be made.
The nonlinear susceptibility is a
connected four-point function for the composite $\bar\psi\psi$ field. The free
fermionic theory is not gaussian in $\bar\psi\psi$, so even in free field
theory
$g_R$ does not vanish. The fact that $g_R\to 0$ near the critical point
indicates that the resulting theory is indeed gaussian in the composite
field which results in a free bosonic theory in the continuum.

The Equation of State (EOS) is obtained from
the effective potential by simple differentiation.
To make the connection with $\delta$, we take $t=0$.
The critical EOS for the magnets is [12]

$$
h\sim {  {M^3}\over{\log(1/M)}  }
\eqno(10)
$$
This defines the exponent $\delta$. Because of the scaling violations,
eq.(10) vanishes {\it faster} then a pure power. So the "effective" $\delta$
is bigger then its mean-field value.
It is convenient to define an effective $\delta$ in the standard way
either as $\tilde\delta =\partial \log h/\partial\log M$ or,
equivalently, through the mass ratio. In any case, the above relation
gives

$$
\tilde\delta=3+{1\over{\log(1/M)}}
\eqno(11)
$$
Thus, on the ratio plot, Fig.2a, the critical isotherm
is no longer flat, but goes down, as the $h$-field (order parameter)
increases. The result of
eq.(11)
is well known in the literature and has been obtained in the past
using the $\epsilon$-expansion: $\delta=3+\epsilon$ [12],
where the correspondence
with eq.(10) is made after the replacement $1/\epsilon\to \log$.
Such corrections to $\delta$, as eq.(11), can never occur in
the case of the chiral transition.

For the four-fermi model the critical EOS [14] reads,

$$
m\sim   <\bar\psi\psi>^3\log(1/<\bar\psi\psi>)
\eqno(12)
$$
Unlike scalar theories, the log's appear in the {\it numerator} --
the right hand side in eq.(12)
vanishes {\it slower} then the pure power and the "effective" $\delta$
is {\it smaller} then the (pure) mean-field value.

$$
\tilde\delta=3-{1\over{\log(1/<\bar\psi\psi>)}}
\eqno(13)
$$
This difference in the position of the logarithm, eqs.(10,12) or, equivalently,
the sign of the scaling violations, eqs.(11,13),
is generic for the two models.
It is independent of the approximation and follows
from the differences in the physics of the two systems.

It is easy to compute logarithmic corrections for other exponents [11].
It turns out that, with these corrections, the hyperscaling relations
become strict inequalities for both systems,
e.g. $d\nu >2-\alpha$, $\gamma/\nu<2-\eta$,
$\beta/\nu <1+\eta/2$, etc.. This statement is equivalent to triviality.
\vskip0.5truecm

\noindent{\bf 4. Discussion and conclusions}
\vskip3truemm

We have seen that the two different realizations of spontaneous symmetry
breaking can be expressed simply in terms of universal quantities.
In this way, many apparently complicated dynamical questions
become transparent. In addition, some of our observations lead to
practical applications --  they can be used for extracting the properties of
the continuum limit of theories with new
fixed points, especially when clear theoretical ideas about
the low-energy physics of the theory are missing. Of special importance is
the knowledge of the
position of the logarithms when triviality is studied on the lattice. It
is extremely difficult to establish the presence of the logarithms for a
finite system and  to disentangle them from finite size
effects. The bounds obtained in this paper establish some criteria in this
direction as far as chiral transitions are concerned.
Recently, they were proven to be decisive
in studies of the chiral transition of $QED$ [15] and
in establishing triviality of the NJL model
in four dimensions by computer simulations [16].

The literature on fermionic $QED$ abounds with loose statements such as
"$QED$ is ultimately trivial and reduces to $\lambda\phi^4$". If $QED$ suffers
from complete screening (the Moscow zero), then we expect the NJL
model to describe its triviality. One result of this paper is that
$\lambda\phi^4$ {\it cannot} be a good guide to fermionic $QED$ under any
circumstances!

There are several physical
implications that the two bounds on $\delta$ imply and we
discuss some of them briefly below.

The wavefunction renormalization constant respects the
Lehmann bound: $0\leq Z\leq 1$. Roughly speaking, $Z$ is the
probability that the scalar field creates a single particle from the vacuum.
The limit $Z=1$ corresponds
to a noninteracting theory whereas the compositeness
condition, $Z=0$, sets an upper bound on the effective coupling [17].
The anomalous dimension
$\eta$ determines the scaling of $Z$ in the critical region
$Z\sim \xi^{-\eta}$. It describes the scaling of the correlation function
in the massless limit: $D(x)\sim 1/|x|^{d-2+\eta}$.
Small $\eta$ is associated with weak coupling
and $\eta=O(1)$ with the strong coupling limit of the theory.
It is related to the exponent $\delta$ through the hyperscaling
relation

$$
\delta= { {d+2-\eta}\over{d-2+\eta} }
\eqno(14)
$$
In the absence of an anomalous dimension, $\delta$ tends to grow above 3 as
the dimensionality is reduced below $d=4$.
In magnets this tendency is respected even for
$\eta\not=0$ as a consequence of the bound $\delta>3$. In all known examples
of magnets with continuous symmetry breaking, $\eta$ remains small suggesting
that the continuum theory is in some sense weakly interacting. For example,
the large-$N$ expansion gives $\eta=O(1/N)$ [6].

The fermion situation is quite different. As a consequence of the
bound $\delta<3$, the tendency of $\delta$ to grow in lower dimensions is
reversed. To achieve this it is necessary that anomalous dimensions are large,
$\eta=O(1)$, in order to counter the $d-2$ factor in the denominator of
eq.(14). For example, in the large-$N$ expansion of four-fermi theories
(below 4 dimensions), $\eta=4-d$ [7]. Thus, fermionic theories are always
strongly interacting because scalars couple strongly to the constituent
fermions. However,
in the context of the $1/N$ expansion, it should be noted that
in  four-fermi theories, mesons don't interact among themselves
at leading order. The $\sigma\pi\pi$ coupling is $O(1/N)$ as for
magnets.

We have outlined in the introduction
the basic difference between the physics of triviality in two models:
in the case
of fermions, triviality is a consequence of having an interaction
at short distances that is too strong; whereas for scalars, it is the weakness
of the
interaction that leads to triviality. This difference is especially visible
from the behavior of the correlation functions in the two theories.
In the scalar theory, where anomalous dimensions are small, the scaling of
the correlation functions is weakly affected at short distances. They behave
almost like free particle propagators, $D(x)\sim 1/|x|^{d-2}$.
For fermions, however, there are nontrivial changes in scaling due to
strongly interacting dynamics at short distances. In four-fermi theory,
for example, at large-$N$ $\eta=4-d$ and $D(x)\sim 1/|x|^2$ irrespective
of $d$.

Another consequence of the difference in the bounds on $\delta$ concerns
the physics in the broken phase. As an effect of spontaneous symmetry breaking
a trilinear coupling is generated and the decay $\sigma\to \pi\pi$ is
a dominant decay mode
in the Goldstone phase. The pole in the $\sigma$ propagator is
buried in the continuum states and the pion state saturates all the
correlation functions. This is especially visible
in lower dimensions and persists even at finite $h$. For fermions, however,
this is not necessarily the case for the simple reason that the $\pi-\sigma$
mass-squared ratio in the broken phase is bounded by $1/\delta>1/3$, away from
the
chiral limit and for the $\sigma\to \pi\pi$ decay to occur the masses
must satisfy $M_\pi^2/M_\sigma^2 \leq 1/4$. Thus, for appropriately chosen
couplings and bare fermion masses, the decay becomes
kinematically forbidden even in the broken phase.

Regarding the usage of perturbation theory, the rules are different for
magnets and fermions.
The applicability of perturbation theory to magnets was
noted long ago [18]. Its success near two and four dimensions
is not surprising. Below four dimensions, $\phi^4$ posesses an infrared fixed
point
at coupling $g_c\sim O(\epsilon)$. This coupling is an upper bound on the
renormalized coupling i.e. $g_R\sim \epsilon$. So, the $\epsilon$-expansion
is in effect renormalized perturbation theory. The critical exponents
receive corrections of the type $\delta=3+O(g_R), \,\, \beta=1/2+O(g_R)$, etc..
Thus, in $\phi^4$ the success of perturbation theory
is a consequence of the
fact that the infrared fixed point moves to the origin as $d\to 4$.
In the non-linear $\sigma$-model the critical coupling is
an ultraviolet fixed point that
moves to the origin as $d\to 2$. The weak
coupling phase is at low-temperatures. Due to the presence
of Goldstone bosons, all the correlation functions are saturated with
massless states and the entire low-temperature phase is massless. Every point
is a critical point in the limit of vanishing magnetic field.
Thus, the low-temperature expansion is an expansion in powers of $T$.
Terms of the form $\exp(-M/T)$
are absent and there is no danger that they will be omitted by using
perturbation theory.
In this way, in principle, the critical region can be accessed through
perturbation theory [18].
Clearly, such reasoning can not be applied to fermions simply because
the weak coupling phase is symmetric. Thus, no matter how small the
coupling is, perturbation theory omits the Goldstone physics as a matter of
principle. It can not produce bound states that accompany the chiral transition
and its applicability is questionable in general. Especially, it is difficult
to imagine how perturbation theory could give a mass ratio  that is constant,
independent of the bare parameters, once the bare coupling is tuned to
the critical value. Even if this were possible, the renormalized coupling
would be sensitive to the variation of the bare mass leading to conflicting
renormalization group trajectories as a consequence.

Finally, we comment on one possible use of the $\delta<3$ bound for controlling
finite size effects in lattice studies of chiral transitions. As was
argued in [8], it is convenient to introduce a plot $\chi_\pi^{-1}$ versus
$<\bar\psi\psi>^2$. The usefulness of this plot becomes clear if we
write the critical EOS. From
$m\sim <\bar\psi\psi>^\delta$ and $\chi_\pi^{-1}=m/<\bar\psi\psi>$, it
follows that

$$
\chi_\pi^{-1}\sim (<\bar\psi\psi>^2)^{(\delta-1)/2}
\eqno(15)
$$
As a consequence of the $\delta<3$ inequality, $(\delta-1)/2 <1$ and the curves
$\chi_\pi^{-1}$ versus $<\bar\psi\psi>^2$ are always concave {\it downwards}.
On a
small lattice the order parameter is smaller and pion mass is bigger then
in the thermodynamic limit. Thus, small volume distortions are always in
the direction of opposite concavity of the plot and the wrong concavity
of this plot is a clear sign of the presence of finite size effects [8].
\vskip0.5truecm

\noindent{\bf Acknowledgement}
We wish to acknowledge the discussions with
E. Fradkin, A. Patrascioiou and E. Seiler.
This work is supported by NSF-PHY 92-00148.
\vfill\eject

\centerline{\bf References}

\noindent
[1] See for example {\it The Standard Model Higgs Boson},
Edited by M. Einhorn, (North-Holland, Amsterdam, 1991).

\noindent
[2] Y. Nambu, in {\it New Trends in Physics}, proceedings of the
XI International Symposium on Elementary Particle Physics,
Kazimierz, Poland, 1988, edited by Z. Ajduk S. Pokorski and A. Trautman (World
Scientific, Singapore, 1989);   V. Miransky, M. Tanabashi and
K. Yamawaki, Mod. Phys. Lett. {\bf A4}, 1043 (1989);
W. Bardeen, C. Hill and M. Lindner, Phys. Rev. {\bf D41}, 1647 (1990).

\noindent
[3] Y. Nambu and G. Jona-Lasinio, Phys. Rev. {\bf 122}, 345 (1961).

\noindent
[4] M. Aizenman, Comm. Math. Phys. {\bf 86}, 1 (1982);
C. Arag\~ ao de Carvalho, S. Caracciolo and J. Fr\" ohlich,
{Nuc. Phys.} {\bf B215}[FS7], 209 (1983).

\noindent
[5] M. L\" uscher and P. Weisz, {Nuc. Phys.} {\bf B290}[FS20], 25 (1987)

\noindent
[6] S. K. Ma, in {\it Phase Transitions
and Critical Phenomena} Vol.6, eds. C. Domb and M. Green (Academic Press,
London, 1976).

\noindent
[7]
S.~Hands, A.~Koci\'{c} and J.~B.~Kogut, Phys. Lett. {\bf B273}
(1991) 111.

\noindent
[8] A. Koci\' c, J. B. Kogut and M.-P. Lombardo,
{Nuc. Phys.} {\bf B398}, 376 (1993).

\noindent
[9] K. Wilson and J. Kogut, Phys. Rep. {\bf 12C}, 75 (1974).

\noindent
[10] See, for example,
C.~Itzykson and J.-M.~Drouffe, Statistical Field Theory (Cambridge
University Press, 1989); V. Privman, P.C. Hohenberg and
A. Aharony, in {\it Phase Transitions
and Critical Phenomena} Vol.14, eds. C. Domb and J.L. Lebowitz
(Academic Press, London, 1991).

\noindent
[11] B. Freedman and G.~A.~Baker Jr, J. Phys. {\bf A15} (1982) L715;
 R.~Schrader, Phys. Rev. {\bf B14} (1976) 172;
B.~D.~Josephson, Proc. Phys. Soc. {\bf 92} (1967) 269, 276.

\noindent
[12]
E. Brezin, J.-C.~Le~Guillou and J.~Zinn-Justin, in {\it Phase Transitions
and Critical Phenomena} Vol.6, eds. C. Domb and M. Green (Academic Press,
London, 1976).

\noindent
[13] T. Eguchi, Phys. Rev. {\bf D17}, 611 (1978).

\noindent
[14] S.~Hands, A.~Koci\'{c} and J.~B.~Kogut, Ann. Phys. {\bf 224}, 29 (1993).

\noindent
[15]  A.~Koci\'{c}, J.~B.~Kogut and K.~C.~Wang, Nucl. Phys. {\bf B398}
(1993) 405.

\noindent
[16] S. Kim, A. Koci\' c and J. Kogut (unpublished)

\noindent
[17] S. Weinberg, Phys. Rev. {\bf 130}, 776 (1963).

\noindent
[18] E. Brezin and J. Zinn-Justin, Phys. Rev. {\bf B14}, 3110 (1976).
\vfill\eject

\noindent{\bf Figure captions}

\noindent
1. Susceptibility ratio as a function of magnetic field (bare mass) for
fixed values of the temperature (coupling).

\noindent
2. The behavior of the critical mass ratio,
$R(t=0,h)=1/\delta$, for different values of $d$, in a)
magnets and b) in the case of chiral transition.

\noindent
3. Critical mass ratio, $R(t=0,h)$, for fermions and magnets
in four dimensions over extended range of magnetic field (mass).

\end